\documentclass[english,aps,tightenlines,showpacs,showkeys,notitlepage]{revtex4-1}
\usepackage[latin9]{inputenc}
\usepackage{amstext}
\usepackage{amssymb}
\usepackage{graphicx}
\usepackage{esint}

\makeatletter

\usepackage{dcolumn}
\usepackage{bm}

\usepackage{babel}

\makeatother

\usepackage{babel}
\begin{document}

\title{Generalized entropic gravity from modified Unruh temperature}

\author{Salih Kibaro\u{g}lu}
\email{salihkibaroglu@gmail.com}

\date{\today}
\begin{abstract}
In this study, the effects of the generalized uncertainty principle
on the theory of gravity are analyzed. Inspired by Verlinde's entropic
gravity approach and using the modified Unruh temperature, the generalized
Einstein field equations with cosmological constant are obtained and
corresponding conservation law is investigated. The resulting conservation
law of energy-momentum tensor dictates that the generalized Einstein
field equations are valid in a very limited range of accelerations.
Moreover, the modified Newton's law of gravity and the modified Poisson
equation are derived. In a certain limit, these modified equations
reduce to their standard forms. 
\end{abstract}

\affiliation{Department of Physics, Kocaeli University, 41380 Kocaeli, Turkey }


\maketitle

\section{Introduction}

In theoretical physics, one of the main studies is to generalize the
theory of gravity, because we have several good reasons. For instance,
merging efforts of the gravity with the standard model and the unification
of quantum mechanics and Einstein's general relativity are still unclear.
Within this scope, there are many methods to cope with these difficulties
such as the supergravity, noncommutative geometry, quantum gravity,
and string theory.

It is interesting to study the generalized theory of gravity from
thermodynamic perspective. There are several theoretical studies on
the thermodynamical origin of gravity. The early studies were done
by Bekenstein and Hawking \cite{bekenstein1973,hawking1974,hawking1975}.
They considered the formulation of the laws of black hole mechanics
taking into account of the thermodynamic point of view. According
to their studies, a relation has occurred between the black hole area
and its entropy as $S_{BH}=A/4l_{p}^{2}$ which is called the Bekenstein-Hawking
(BH) entropy. Their studies opened a new window to describe the emergent
nature of space-time and gravity. In 1995, Jacobson was the first
who claimed that the gravity may not be a fundamental interaction,
but it is originated from the first law of thermodynamics on a local
Rindler horizon. From this idea, he found certain components of Einstein's
field equation \cite{jacobson1995}. Yet another important work was
achieved by Verlinde \cite{verlinde2011} (see also \cite{verlinde2017})
who formulated a thermodynamical description of gravity by considering
the holographic principle and BH entropy. In this way he obtained
Newton's law of gravity and Einstein's field equations. Similar to
Jacobson's idea, Verlinde suggested that gravity is not a fundamental
force, but emerges as an entropic force. This idea is known as the
entropic gravity or emergent gravity.

The studies in the entropic gravity proposal can be divided into two
approaches \cite{carroll2016}. The first one is the thermodynamic
gravity (TG) which is used to obtain the Einstein field equation from
the BH entropy \cite{jacobson1995,verlinde2011,verlinde2017,cai2005,Padmanabhan2010,tian2010},
and the second one is the holographic gravity (HG) in which the Einstein's
equation can be obtained by keeping entropy stationary in equilibrium
under the variations of space-time geometry and the quantum states
within a small region \cite{faulkner2014,lashkari2014,jacobson2016}.

In the light of the above theoretical background, it is easy to say
that the entropic gravity approach provides a powerful framework to
generalize the theory of gravity. For instance, if we use generalized
thermodynamic quantities such as entropy in the entropic gravity approach,
we can find a generalized model of the theory of gravity \cite{cai2010,ho2010,sheykhi2012,hendi2012,wang2014,moradpour2015,dil2015,senay2018}.
Therefore, it may help us to find a deep understanding of the nature
of space-time and gravity. This idea motivated us to study the relation
between the generalized uncertainty principle (GUP) and entropic gravity.
For this purpose, we consider the GUP corrected Unruh temperature
to modify the theory of gravitation. Therefore, using Verlinde's entropic
gravity approach \cite{verlinde2011} and the modified Unruh temperature
\cite{scardigli2018}, we derive the generalized Einstein field equations
with the cosmological term and analyze corresponding conservation
law. We also find GUP corrected Newton's law of gravity and Poisson's
equation.

The paper is organized as follows. In Sec. 2, the historical background
for the Unruh effect and the modification of Unruh temperature based
on GUP are shortly reviewed. In Sec. 3, the generalized Einstein's
field equations are obtained. In Sec. 4, we find the generalized forms
of Newton's law of gravity and Poisson's equation. The last section
is devoted to the discussion of the obtained results.

\section{Gup corrected Unruh temperature}

Discovery of the Hawking effect \cite{hawking1974,hawking1975} demonstrates
that black holes should radiate with a temperature $T_{H}=\hbar g/2\pi k_{B}c$
where $g$ is the gravitational acceleration, $k_{B}$ is Boltzman's
constant and $c$ is the speed of light. The Hawking effect also describes
the black-body radiation which is produced by the vacuum fluctuations
near the event horizon of the black hole. Moreover, there is another
approach which is so-called the Unruh effect (or the Fulling-Davies-Unruh
effect) \cite{fulling1973,davies1975,unruh1976}. The Unruh effect
describes the motion of an accelerated observer in a flat Minkowski
spacetime. In this case, the observer observes black-body radiation
released from the Rindler horizon and experiences following temperature,

\begin{equation}
T_{U}=\frac{\hbar a}{2\pi k_{B}c},\label{eq: temp unruh}
\end{equation}
as though it is in a thermal field where an inertial observer would
observe none. Here, $a$ is the acceleration of the reference frame.
This temperature is called the Unruh temperature (or the Hawking-Unruh
temperature). The Unruh effect, similar to the Hawking effect, is
originated from the vacuum fluctuations which cause a particle-antiparticle
pairs creation near the Rindler horizon of the accelerated frame.
This effect is important to understand the notion of particle emission
from cosmological horizons and black holes (for more detail see \cite{alsing2004,crispino2008}).

The formulation of the Unruh effect can be obtained by using the Heisenberg
uncertainty principle (HUP) when we take into account of a photon
that has crossed the Rindler event horizon. In this case, the position
uncertainty of the photon depends on the crossing point of the Rindler
event horizon. In this framework, a direct derivation of the Unruh
effect with the help of HUP can be found in \cite{scardigli1995}
(see also \cite{gine2018}).

We know from early studies \cite{snyder1947,yang1947,mead1964,karolyhazy1966,amati1987,amati1988,gross1988,maggiore1993,Ng1994,kempf1995,Rovelli1996,scardigli1999,adler1999}
that HUP should be generalized when we consider the Planck scale where
the gravitational effect is neglected. So, we can say that such a
generalization may lead to modify the Unruh effect. In this case,
we can use the generalized uncertainty principle (GUP) which provides
the determination of the quantum gravitational corrections for HUP.
This approximation is used in various branches of physics such as
the string theory, loop quantum gravity, and non-commutative quantum
mechanics (for more detail see review \cite{tawfik2015}). For instance,
GUP is used to find quantum gravitational corrections for the Planck-scale
black hole thermodynamics including the Unruh temperature \cite{scardigli2018}
and the Hawking temperature \cite{scardigli2017}. Now, we shall briefly
review the paper \cite{scardigli2018} which includes the GUP corrected
Unruh temperature. In order to obtain a generalized version of the
Unruh temperature, the authors used the following form of GUP, 
\begin{equation}
\Delta x\Delta p\geq\frac{\hbar}{2}\left[1+\beta\left(\frac{\Delta p}{m_{p}}\right)^{2}\right],\,\,\,\,\,\,\,\,\,\,\left[\hat{x},\hat{p}\right]=i\hbar\left[1+\beta\left(\frac{\hat{p}}{m_{p}}\right)^{2}\right],\label{eq: GUP commutation}
\end{equation}
where the constant $\beta$ is a dimensionless deformation parameter
and the other constants are chosen as $c=1$, $k_{B}=1$ throughout
this paper. Considering this GUP background and using the quantum
field theoretic calculations, the generalized Unruh temperature can
be described to the first approximation in $\beta$,
\begin{equation}
T\backsimeq T_{U}\left[1+\beta\pi\Omega\left(\frac{l_{p}a}{\pi}\right)^{2}\right],\label{eq: temp1}
\end{equation}
where $\Omega$ is the Rindler frequency and $l_{p}=\sqrt{\hbar G/c^{3}}\simeq10^{-35}\ \text{m}$
is the Planck length. In a certain condition, we can take the Rindler
frequency as $\Omega\approx1/2\pi$, and then Eq.(\ref{eq: temp1})
takes following form \cite{scardigli2018},

\begin{equation}
T\backsimeq T_{U}\left[1+\frac{\beta}{2}\left(\frac{l_{p}a}{\pi}\right)^{2}\right]=T_{U}\left[1+\frac{\beta}{2}\left(\frac{T_{U}}{m_{p}}\right)^{2}\right].\label{eq: temp2}
\end{equation}
Thus, taking account of the Eq.(\ref{eq: temp unruh}), the temperature
$T$ is obtained with a dependence of the cubic power of the acceleration.
We can also write the last equations in a simple form,

\begin{equation}
T=T_{U}f\left(a\right)=\frac{\hbar a}{2\pi}f\left(a\right),\label{eq: mod temp}
\end{equation}
where the function $f\left(a\right)$ is defined by,

\begin{equation}
f\left(a\right)=\left[1+\frac{\beta}{2}\left(\frac{l_{p}a}{\pi}\right)^{2}\right]=\left[1+\frac{\beta}{2}\left(\frac{T_{U}}{m_{p}}\right)^{2}\right],\label{eq: f(a)}
\end{equation}
where $m_{p}=\hbar/2l_{p}\simeq10^{-8}\ \text{kg}$ is the Planck
mass. This correction function tends to increase the Unruh temperature
for selected GUP model. Different kind of $f\left(a\right)$ functions
can also be found in the literature (for the maximal acceleration
principle \cite{benedetto2015,luciano2019} or the extended uncertainty
principle \cite{chung2019}). Setting the deformation parameter as
$\beta=0$, it is easy to see that Eq.(\ref{eq: temp2}) reduces to
the standard Unruh temperature.

\section{Generalized Einstein field equations }

The thermodynamic origin of gravity, also known as the entropic gravity,
is a theory that describes the gravitational force by using thermodynamic
quantities such as energy, temperature, and entropy. Thus, if one
modifies the mentioned quantities, the gravity should be modified.
In the study \cite{wang2014}, Wang unified several modified entropic
gravity models by inserting a model dependent factor $f\left(a,A\right)$
into the equipartition rule $E=\frac{aA}{4\pi G}f\left(a,A\right)$.
In this idea, he found modified Einstein's and Newton's gravitational
field equation by considering $E$ as total energy in a closed holographic
surface. Here, $a$ corresponds to the (redshifted) surface acceleration
and $A$ is the area of the holographic surface. In this section,
we will follow Verlinde's entropic gravity proposal \cite{verlinde2011}
and Wang's method to find quantum corrected Einstein's field equations
by using the GUP corrected Unruh temperature in Eq.(\ref{eq: mod temp}).
In analogy to Wang's study, we will use $f\left(a\right)$ in Eq.(\ref{eq: f(a)})
as the model dependent correction factor.

\begin{figure}
\includegraphics[scale=0.8]{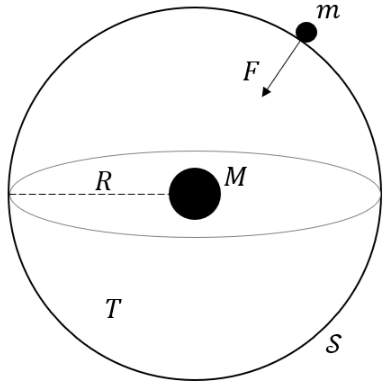}\caption{Suppose that we have two masses, one is a test particle with mass
$m$ close to spherical holographic screen $\mathcal{S}$, and the
other is the source with mass $M$ surrounded by the screen.\label{fig:Suppose-that-we}}
\end{figure}

Verlinde considers the black hole horizon as a spherically symmetric
holographic screen $\mathcal{S}$ based on Bekenstein's idea \cite{bekenstein1973}.
In this consideration, we have a test particle with mass $m$ and
an acceleration $a$ moving along to the screen (see Figure \ref{fig:Suppose-that-we}).
To find the relativistic gravitational field equation, we consider
a static background with a global time-like Killing vector $\xi^{a}$
which is normalized as $\xi^{a}\xi_{a}=-e^{2\phi}$. Here, $\phi$
represents the natural generalization of Newton's potential in general
relativity \cite{wald1984} which can be written as

\begin{equation}
\phi=\frac{1}{2}\ln\left(-\xi^{a}\xi_{a}\right).\label{eq: newton pot}
\end{equation}
We also note that the exponential of Eq.(\ref{eq: newton pot}) represents
the redshift factor that relates the local time coordinate at a reference
point with $\phi=0$.

In our case, we will use $\phi$ to describe a foliation of space
and consider our holographic screen at surfaces of constant redshift.
Because the entire screen uses the same time coordinate (for more
details see \cite{verlinde2011}). Let us consider a force that acts
on the particle of mass $m$. To obtain the emergence of inertia and
the equivalence principle, we have to relate this Killing vector field
with the temperature and the entropy gradients. In this framework,
we want to show that the usual geodesic motion of particles can be
considered as being the result of an entropic force. The four-velocity
$u^{b}$ and its four-acceleration $a^{b}$ of the particle can be
expressed by using the Killing vector field as,

\begin{equation}
u^{b}=e^{-\phi}\xi^{b},\,\,\,\,\,\,\,\,\,\,\,a^{b}=e^{-2\phi}\xi^{a}\nabla_{a}\xi^{b}.
\end{equation}
With the help of the Killing equation $\nabla_{a}\xi_{b}+\nabla_{b}\xi_{a}=0$
and Eq.(\ref{eq: newton pot}), the four-acceleration can be defined
in terms of the potential $\phi$ as,

\begin{equation}
a^{b}=-\nabla^{b}\phi.\label{eq: acceleration}
\end{equation}
Here, the potential $\phi$ is just used as a device which satisfies
the relation in Eq.(\ref{eq: newton pot}) to define the acceleration.
Moreover, the acceleration is perpendicular to the screen, so we can
write the acceleration as a scalar quantity by contracting it with
a unit outward pointing vector $N^{a}$ which is normal to both the
holographic surface $\mathcal{S}$ and the Killing vector $\xi^{a}$.
From these definitions, the local temperature $T$ on the holographic
screen is defined in analogy to Eq.(\ref{eq: mod temp}),

\begin{equation}
T=\frac{\hbar}{2\pi}f\left(\nabla\phi\right)e^{\phi}N^{a}\nabla_{a}\phi,\label{eq: temp}
\end{equation}
where the GUP correction function transformed as $f\left(a\right)\rightarrow f\left(\nabla\phi\right)$
due to the definition of the acceleration in Eq.(\ref{eq: acceleration}).
Here, we used the exponential $e^{\phi}$ as a redshift factor, because
the temperature is measured with respect to the reference point at
infinity \cite{verlinde2011}.

To find the force that acts on a particle which is very close to the
screen, we first consider the change in entropy at the screen is $2\pi$
for a displacement by one Compton wavelength normal to the screen,

\begin{equation}
\nabla_{a}S=-2\pi\frac{m}{\hbar}N_{a}.
\end{equation}
So, one can obtain the entropic force by using the last equation and
Eq.(\ref{eq: temp}) as follows,

\begin{equation}
F_{a}=T\nabla_{a}S=-me^{\phi}f\left(\nabla\phi\right)\nabla_{a}\phi.\label{eq: rel entropic force}
\end{equation}
This is the GUP corrected model of the relativistic analogue of Newton's
second law of gravity. It defines the gravitational force that acts
on the particle which located close to the screen $\mathcal{S}$.
If we take the limit of low acceleration and low potential where the
redshift factor can be ignored, then Eq.(\ref{eq: rel entropic force})
reduces its non-relativistic form $F=maf\left(a\right)$.

Now, we would like to find a generalized version of Einstein's field
equation by combining Verlinde's entropic gravity proposal with the
GUP model in Eq.(\ref{eq: GUP commutation}). According to Bekenstein,
if there is a test particle which closes to the black hole horizon
a one Compton wavelength, the test particle increases the mass and
horizon area of the black hole. This process is defined as one bit
of information. Considering the holographic screen on a closed surface
with constant redshift, the total number of bits $N$ on the screen
can be specified as,

\begin{equation}
N=\frac{A}{G\hbar},\label{eq: bit}
\end{equation}
where $A$ is the area of closed surface on the screen and $G$ is
a constant which will be identified as the Newton's gravitational
constant. In this case, we assume that the energy which is related
to source mass $M$ is distributed over all the bits. According to
the equipartition law of energy $E=\frac{1}{2}TN$, and using the
relation $E=M$, the total mass can be written as

\begin{eqnarray}
M & = & \frac{1}{2}\oint_{\mathcal{S}}TdN.\label{eq: mass}
\end{eqnarray}
Substituting the Eq.(\ref{eq: temp}) and Eq.(\ref{eq: bit}) in Eq.(\ref{eq: mass}),
we get,

\begin{eqnarray}
M & = & \frac{1}{4\pi G}\oint_{\mathcal{S}}f\left(\nabla\phi\right)e^{\phi}\nabla\phi dA.\label{eq: mass2}
\end{eqnarray}
This equation can be seen as the natural generalization of Gauss'
law to the General Relativity and $M$ corresponds to the Komar mass.
Here, the differential surface element is described by, 
\begin{equation}
\left|dA\right|=dx^{a}\wedge dx^{b}\epsilon_{abcd}e^{-\phi}\xi^{c}N^{d}.
\end{equation}
Thus, the Komar mass can be written as follows by using the Stokes'
theorem, Eq.(\ref{eq: newton pot}) and $\nabla_{a}\nabla^{a}\xi^{b}=-R_{\,\,\,a}^{b}\xi^{a}$
\cite{sheykhi2012,wang2014}, 

\begin{equation}
M=\frac{1}{4\pi G}\intop_{\mathcal{V}}\left\{ R_{ab}\xi^{b}n^{a}+\left[\nabla_{b}f\left(a\right)\right]\left[\nabla_{a}\xi^{b}\right]n^{a}\right\} dV,\label{eq: mass6}
\end{equation}
where $R_{ab}$ is the Ricci curvature tensor, $\mathcal{V}$ is the
three dimensional volume bounded by the holographic screen and $n^{a}$
is its outward normal. For simplicity, we used $f\left(a\right)$
rather than $f\left(\nabla\phi\right)$. Due to the relation $\xi^{b}\nabla_{b}f\left(a\right)=0$
which is demonstrated in \cite{wang2014}, the second term in the
parenthesis can be written by, 
\begin{equation}
\left[\nabla_{b}f\left(a\right)\right]\left[\nabla_{a}\xi^{b}\right]n^{a}=-n^{a}\xi^{b}\nabla_{a}\nabla_{b}f\left(a\right).
\end{equation}
The Komar mass can also be written as an integral over the enclosed
volume of certain components of the energy-momentum tensor $T_{ab}$
\cite{wald1984} as, 
\begin{eqnarray}
M & = & 2\int_{\mathcal{V}}\left(T_{ab}-\frac{1}{2}g_{ab}T\right)\xi^{b}n^{a}dV,\label{eq: mass7}
\end{eqnarray}
where $g_{ab}$ is the spacetime metric tensor. If we compare Eqs.
(\ref{eq: mass6}) and (\ref{eq: mass7}), we obtain the generalized
Einstein field equations as follows,

\begin{equation}
f\left(a\right)R_{ab}-\nabla_{a}\nabla_{b}f\left(a\right)=8\pi G\left(T_{ab}-\frac{1}{2}g_{ab}T\right).\label{eq: einstein1}
\end{equation}
This is only a time-time component of the field equation \cite{verlinde2011}.
The last expression can also be written as,

\begin{equation}
R_{ab}-\frac{1}{2}g_{ab}R-\frac{1}{f\left(a\right)}\left(\nabla_{a}\nabla_{b}-\frac{1}{2}g_{ab}\nabla^{2}\right)f\left(a\right)=\frac{8\pi G}{f\left(a\right)}T_{ab},\label{eq: einstein2}
\end{equation}
and in a simple form,

\begin{equation}
R_{ab}-\frac{1}{2}g_{ab}R=\frac{8\pi G}{f\left(a\right)}\left\{ T_{ab}+T_{ab}\left(a\right)\right\} .\label{eq: einstein2-1}
\end{equation}
The last result can be seen as the GUP corrected Einstein field equations
with a new additional energy-momentum tensor as a function of the
acceleration which is defined, 
\begin{equation}
T_{ab}\left(a\right)=\frac{1}{8\pi G}\left(\nabla_{a}\nabla_{b}f\left(a\right)-\frac{1}{2}g_{ab}\nabla^{2}f\left(a\right)\right).
\end{equation}
Due to this new contribution, the energy momentum conservation law
should be revisited. Considering the usual form of Einstein's field
equations $G_{\,\,b}^{a}=8\pi GT_{\,\,b}^{a}$, the covariant conservation
law of the energy-momentum tensor can be written as $\nabla_{a}T_{\,\,b}^{a}=0$,
and in accordance with this result the covariant derivative of the
Einstein tensor $G_{\,\,b}^{a}$ goes to zero,

\begin{equation}
\nabla_{a}\left(R_{\,\,b}^{a}-\frac{1}{2}\delta_{\,\,b}^{a}R\right)=0.\label{eq: R cons law}
\end{equation}

In our case, the gravitational field equation in Eq.(\ref{eq: einstein2})
is more complicated than the standard one. In order to determine corresponding
conservation law, we take the covariant derivative of Eq.(\ref{eq: einstein2}),
then we get the following equations,

\begin{equation}
R\nabla_{a}f\left(a\right)+\nabla_{a}\nabla^{2}f\left(a\right)=-16\pi G\nabla_{a}T_{\,\,b}^{a}.
\end{equation}
The right-hand side of the last equation should be zero because of
the well-known relation $\nabla_{a}T_{\,\,b}^{a}=0$, so we find following
constraint, 
\begin{equation}
R\nabla_{a}f\left(a\right)=-\nabla_{a}\nabla^{2}f\left(a\right).\label{eq: constraint}
\end{equation}
As a consequence of the last equation, one can say that the energy
momentum tensor is conserved only if the constraint is satisfied.
In other words, it can be seen as a restriction for this kind of modified
entropic gravity because only few values for $a^{b}=-\nabla^{b}\phi$
are permitted by this requirement. According to Wang's paper \cite{wang2014},
this result reminds the Chern-Simons (CS) modification of general
relativity \cite{jackiw2003} in which conservation of the energy-momentum
is unwarranted by gravitational equations. So, one can consider the
last equation as a criterion of admissible metrics in modified models
of entropic gravity. In the limit of $\beta=0$, we therefore recover
the standard Einstein field equation,

\begin{equation}
R_{ab}-\frac{1}{2}g_{ab}R=8\pi GT_{ab}.
\end{equation}
Yet another alternative is to consider Komar mass formula with nonzero
cosmological constant $\Lambda$ \cite{ho2010,wald1984},

\begin{equation}
M=2\int_{\mathcal{V}}\left(T_{ab}-\frac{1}{2}g_{ab}T+\frac{\Lambda}{8\pi G}g_{ab}\right)n^{a}\xi^{b}dV.
\end{equation}
If we compare the last equation with Eq.(\ref{eq: mass6}), we get
the following field equations, 
\begin{equation}
R_{ab}-\frac{1}{2}g_{ab}R+\frac{\Lambda}{f\left(a\right)}g_{ab}=8\pi\frac{G}{f\left(a\right)}\left\{ T_{ab}+T_{ab}\left(a\right)\right\} ,\label{eq: einstein2-1-1}
\end{equation}
which is the generalized Einstein field equation with scaled cosmological
constant. In addition, according to Eq.(\ref{eq: einstein2-1-1}),
the constraint equation in Eq.(\ref{eq: constraint}) remains unchanged
because the covariant derivative of cosmological constant is zero.

\section{Modified Newton's law of gravity and Poisson's equation}

\begin{figure}
\includegraphics[scale=0.7]{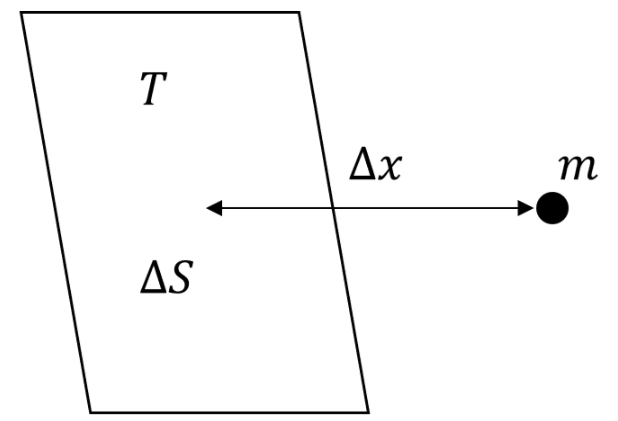}\caption{A particle with mass $m$ is moving to a part of the holographic surface.
The screen which corresponds to emerged part of space contains the
particle and stores data which describes the part of the space that
has not yet emerged \cite{verlinde2011}. \label{fig:A-particle-with}}
\end{figure}

Suppose that a test particle of mass $m$ moves a distance $\Delta x$
orthogonal to the holographic screen as in Figure \ref{fig:A-particle-with}.
In this case, there is a relation between the change of entropy $\Delta S$
and the distance of the particle from screen $\Delta x$ \cite{verlinde2011},

\begin{equation}
\Delta S=2\pi\frac{m}{\hbar}\Delta x.\label{eq: delta S delta x}
\end{equation}
The last equation demonstrates that the entropy is proportional to
the information of the test particle and it can be seen as one of
the main formulas for construction of the entropic force. The change
of entropy given in Eq.(\ref{eq: delta S delta x}) can also be written
as $\Delta S=2\pi\Delta x/\lambda_{m}$ by using the definition of
the Compton wavelength $\lambda_{m}=\hbar/m$. In the case of $\Delta x\simeq\lambda_{m}$,
the particle is absorbed by the screen and this event leads to increase
the entropy of the system \cite{verlinde2011}. According to the Verlinde's
proposal, the entropic force is defined as follows,

\begin{equation}
F=T\frac{\Delta S}{\Delta x},\label{eq: entropic force}
\end{equation}
where $T=T_{0}f\left(a\right)=\frac{\hbar a}{2\pi}f\left(a\right)$
represents the temperature which is defined in analogue to Eq.(\ref{eq: mod temp}).
Here, $a$ represents the acceleration of the particle with mass $m$.
Substituting the temperature and Eq.(\ref{eq: delta S delta x}) in
Eq.(\ref{eq: entropic force}), we get the gravitational force acting
on a particle of mass $m$ as,

\begin{eqnarray}
F & = & maf\left(a\right)=ma\left[1+\frac{\beta}{2}\left(\frac{l_{p}a}{\pi}\right)^{2}\right],\label{eq: newton law non rel}
\end{eqnarray}
or it can alternatively be written in terms of $T_{0}$ by using Eq.(\ref{eq: f(a)}),
\begin{equation}
F=ma\left[1+\frac{\beta}{2}\left(\frac{T_{0}}{m_{p}}\right)^{2}\right].\label{eq: newton law non rel Tu}
\end{equation}
The Eqs. (\ref{eq: newton law non rel}) and (\ref{eq: newton law non rel Tu})
represent the modification of Newton's second law of gravity, $F=ma$,
with the Planck scale correction based on the GUP model in Eq.(\ref{eq: GUP commutation}).
This result is consistent with the non-relativistic analogue to Eq.(\ref{eq: rel entropic force})
which is found in the previous section. Similar modifications can
be found in \cite{gosh2010,nozari2012,santos2012,moradpour2018A,bagchi2019}.

Now we want to determine the effects of the GUP model on Poisson's
equation. For this purpose, we use a holographic screen $\mathcal{S}$
which corresponds to an equipotential surface with fixed Newton potential
$\Phi_{0}$. We will assume that the volume which enclosed by the
screen contains the mass distribution given by $\rho\left(\vec{r}\right)$,
and all test particles are outside of this volume. To define the force
on the particles, we need to identify the temperature \cite{verlinde2011}.
Similar to the Unruh temperature in Eq.(\ref{eq: temp unruh}), we
first define the acceleration. This acceleration can be found by considering
a test particle and moving it close to the screen. The local acceleration
of the test particle can be defined by a potential $\Phi$ as,

\begin{equation}
\vec{a}=-\vec{\nabla}\Phi,
\end{equation}
where the potential $\Phi$ is just used to a variable to describe
the local acceleration, but we don't know yet it has a relation with
the mass distribution. In this condition, the correction factor can
be viewed as $f\left(a\right)\rightarrow f\left(\vec{\nabla}\Phi\right)$.
So, the local temperature $T$ in analogue with Eq.(\ref{eq: mod temp})
can be defined as

\begin{equation}
T=\frac{\hbar}{2\pi}f\left(\vec{\nabla}\Phi\right)\left|\vec{\nabla}\Phi\right|.\label{eq: temp-1}
\end{equation}
Substituting the last equation into Eq.(\ref{eq: mass}) and using
Eq.(\ref{eq: bit}) we get,

\begin{equation}
M=\frac{1}{4\pi G}\oint_{\mathcal{S}}f\left(\vec{\nabla}\Phi\right)\vec{\nabla}\Phi\cdot d\vec{A}.\label{eq: mass_2}
\end{equation}
This result can be written by using the divergence theorem as follows,

\begin{equation}
M=\frac{1}{4\pi G}\intop_{\mathcal{V}}\vec{\nabla}\cdot\left[f\left(\vec{\nabla}\Phi\right)\vec{\nabla}\Phi\right]dV,\label{eq: mass3}
\end{equation}
where $\mathcal{V}$ represents three dimensional volume element.
The mass distribution of the volume in the closed surface $\mathcal{S}$
can also be given as

\begin{equation}
M=\intop_{\mathcal{V}}\rho\left(\vec{r}\right)dV.\label{eq: mass4}
\end{equation}
Comparing Eq.(\ref{eq: mass3}) with Eq.(\ref{eq: mass4}), we get
following equation which contains a relation between the potential
and the mass distribution,

\begin{equation}
\vec{\nabla}\cdot\left[f\left(\vec{\nabla}\Phi\right)\vec{\nabla}\Phi\right]=4\pi G\rho\left(\vec{r}\right).\label{eq: poisson1-1}
\end{equation}
If one takes $f\left(\vec{\nabla}\Phi\right)=1$, then the potential
$\Phi$ satisfies the usual Poisson's equation as follows

\begin{equation}
\nabla^{2}\Phi=4\pi G\rho\left(\vec{r}\right).\label{eq: poisson1}
\end{equation}
Therefore, Eq.(\ref{eq: poisson1-1}) can be interpreted as the quantum
corrected Poisson's equation and this result shows that there is an
exact contribution to the Poisson equation comes from GUP (for similar
modification of Poisson's equation see \cite{sheykhi2012,pikhitsa2010,zhe2011}).

\section{Conclusion}

In this work, we investigated the possible effects of generalized
uncertainty principle (GUP) on Einstein's theory of general relativity
and Newton's law of gravity. For this aim, we considered Verlinde's
entropic gravity proposal and the GUP corrected Unruh temperature.
Using these approaches, we derived generalized version of the Einstein
field equations with scaled cosmological constant and an additional
energy-momentum tensor $T_{ab}\left(a\right)$ by using Wang's perspective
\cite{wang2014}. The additional energy-momentum tensor in Eq.(\ref{eq: einstein2-1-1})
dictates that the modified temperature gives rise to a new source
term in Einstein's field equation. This kind of source term may be
related to the dark energy \cite{frieman2008,padmanabhan2009}. For
this reason, the cosmological consequences of the Eq.(\ref{eq: einstein2-1})
and Eq.(\ref{eq: einstein2-1-1}) are open problems. Besides, if one
consider the $f\left(R\right)$ theories of gravity (for more detail
see review \cite{felice2010}), the modified Einstein's field equation
can be given as,
\begin{equation}
R_{ab}-\frac{1}{2}\left[\frac{f\left(R\right)}{\partial_{R}f\left(R\right)}\right]g_{ab}-\frac{1}{\partial_{R}f\left(R\right)}\left(\nabla_{a}\nabla_{b}-g_{ab}\nabla^{2}\right)\partial_{R}f\left(R\right)=\frac{8\pi G}{\partial_{R}f\left(R\right)}T_{ab},
\end{equation}
where $f\left(R\right)$ is a function of the Ricci scalar $R$, $\partial_{R}$
is the derivative with respect to $R$ and $T_{ab}$ is the energy-momentum
tensor of the matter fields. In our context, if one take the function
as $f\left(R\right)=f\left(a\right)R$ by the help of Eq.(\ref{eq: f(a)}),
there is an exact similarity arises between the last equation and
the GUP corrected Einstein's field equation in Eq.(\ref{eq: einstein2}).
In addition, it is well-known that the equivalence between the $f\left(R\right)$
theory and the scalar-tensor theories, so one may find different relations
between $f\left(R\right)$ and $f\left(a\right)$ in the framework
of the Brans-Dicke theory or the conformal gravity.

We also analyzed the energy-momentum conservation law of the generalized
Einstein field equations and found a constraint Eq.(\ref{eq: constraint})
which is required to satisfy corresponding conservation law. As a
consequence of this constraint, according to Wang \cite{wang2014},
the modifications of Einstein's equation in Eq.(\ref{eq: einstein2-1-1})
are valid only in very restricted conditions since only a few values
for $a^{b}$ are allowed by this constraint.

The relativistic and nonrelativistic forms of Newton's law of gravity
were obtained with GUP corrections in Eqs. (\ref{eq: rel entropic force})
and (\ref{eq: newton law non rel}) respectively. These modifications
have similar prediction with the Randall-Sundrum II model \cite{randall1999}
which includes one uncompactified extra dimension. Parallel results
for different models of GUP can be found in \cite{awad2014}. Also,
the GUP corrected Poisson equation was found by using the entropic
gravity approximation in Eq.(\ref{eq: poisson1-1}).

The deformation parameter $\beta$ was studied for the gravitational
and non-gravitational regimes in the literature (the corresponding
results was reviewed in \cite{lambiase2018}). For the gravitational
regime, the deformation parameter should be $\beta<10^{69}$ for the
perihelion precession (Solar system data) in gravitational measurements.
According to the paper \cite{scardigli2017}, the parameter was also
calculated as $\beta=82\pi/5$ in the context of the quantum field
theory and the general theory of relativity. Taking the consideration
of the non-gravitational case, the parameter takes values of $\beta<10^{20}$
and $\beta<10^{46}$, respectively, for the Lamb shift and the Landau
levels. Our results may provide a new background to study the mentioned
areas \cite{lambiase2018} for the gravitational regime. Besides,
it is easy to show that, in a certain limit ($\beta=0$), the results
reduce to their conventional forms.

Our results and the papers \cite{cai2010,ho2010,sheykhi2012,hendi2012,wang2014,moradpour2015,dil2015,senay2018}
show that the entropic gravity proposal provides a powerful framework
to obtain the generalized theory of gravity by using thermodynamical
quantities. 
\begin{acknowledgments}
The author would like to thank Oktay Cebecio\u{g}lu and Mustafa Senay
for valuable discussions. 
\end{acknowledgments}


\begin{thebibliography}{10}
\bibitem{bekenstein1973} J. D. Bekenstein, Phys. Rev. D \textbf{7},
2333 (1973).

\bibitem{hawking1974} S. W. Hawking, Nature \textbf{248}, 30 (1974).

\bibitem{hawking1975} S. W. Hawking, Commun. Math. Phys. \textbf{43},
199 (1975).

\bibitem{jacobson1995} T. Jacobson, Phys. Rev. Lett. \textbf{75},
1260 (1995).

\bibitem{verlinde2011} E. Verlinde, J. High Energy Phys. \textbf{04},
029 (2011).

\bibitem{verlinde2017} E. Verlinde, SciPost Phys. \textbf{2}, 016
(2017).

\bibitem{carroll2016} S. M. Carroll, G. N. Remmen, Phys. Rev. D \textbf{93},
124052 (2016).

\bibitem{cai2005} R. G. Cai, S. P. Kim, JHEP \textbf{02}, 050 (2005).

\bibitem{Padmanabhan2010} T. Padmanabhan, Rept. Prog. Phys. \textbf{73},
046901 (2010).

\bibitem{tian2010} Y. Tian, X. N. Wu, Phys. Rev. D \textbf{81}, 104013
(2010).

\bibitem{faulkner2014} T. Faulkner, M. Guica, T. Hartman, R. C. Myers,
M. van Raamsdonk, JHEP \textbf{1403}, 051 (2014).

\bibitem{lashkari2014} N. Lashkari, M. B. McDermott, M. Van Raamsdonk,
JHEP \textbf{04}, 195 (2014).

\bibitem{jacobson2016} T. Jacobson, Phys. Rev. Lett. \textbf{116},
201101 (2016).

\bibitem{cai2010} R. G. Cai, L. M. Cao, N. Ohta, Phys. Rev. D \textbf{81},
061501(R) (2010).

\bibitem{ho2010} C. M. Ho, D. Minic, Y. J. Ng, Phys. Lett. B \textbf{693},
567 (2010).

\bibitem{sheykhi2012} A. Sheykhi, S. K. Rezazadeh, JCAP \textbf{1210},
012 (2012).

\bibitem{hendi2012} S. H. Hendi, A. Sheykhi, Int. J. Theor. Phys.
\textbf{51}, 1125 (2012).

\bibitem{wang2014} T. Wang, Sci. China-Phys. Mech. Astron., \textbf{57},
1623-1629 (2014).

\bibitem{moradpour2015} H. Moradpour, A. Sheykhi, Int. J. Theor.
Phys. \textbf{55}, 1145-1155 (2016).

\bibitem{dil2015} E. Dil, Can. J. Phys. \textbf{93}, 1274 (2015).

\bibitem{senay2018} M. Senay, S. Kibaro\u{g}lu, Int. J. Mod. Phys.
A \textbf{33}, 1850218 (2018).

\bibitem{scardigli2018} F. Scardigli, M. Blasone, G Luciano, R. Casadio,
Eur. Phys. J. C \textbf{78}, 728 (2018).

\bibitem{fulling1973} S. A. Fulling, Phys. Rev. D \textbf{7}, 2850
(1973).

\bibitem{davies1975} P. C. W. Davies, J. Phys. A \textbf{8}, 609
(1975).

\bibitem{unruh1976} W. G. Unruh, Phys. Rev. D \textbf{14}, 870 (1976).

\bibitem{alsing2004} P. M. Alsing, P. W. Milonni, American Journal
of Physics \textbf{72}, 1524 (2004).

\bibitem{crispino2008} L. C. B. Crispino, A. Higuchi, G. E. A. Matsas,
Rev. Mod. Phys. \textbf{80}, 787 (2008).

\bibitem{scardigli1995} F. Scardigli, Nuovo Cim. B \textbf{110},
1029 (1995).

\bibitem{gine2018} J. Giné, EPL \textbf{121}, 10001 (2018).

\bibitem{snyder1947} H. S. Snyder, Phys. Rev. \textbf{71}, 38 (1947).

\bibitem{yang1947} C. N. Yang, Phys. Rev. \textbf{72}, 874 (1947).

\bibitem{mead1964} C. A. Mead, Phys. Rev. \textbf{135}, B849 (1964).

\bibitem{karolyhazy1966} F. Karolyhazy, Nuovo Cim. A \textbf{42},
390 (1966).

\bibitem{amati1987} D. Amati, M. Ciafaloni, G. Veneziano, Phys. Lett.
B \textbf{197, }81 (1987).

\bibitem{amati1988} D. Amati, M. Ciafaloni, G. Veneziano, Int. J.
Mod. Phys. A \textbf{3}, 1615 (1988).

\bibitem{gross1988} D. J. Gross, P. F. Mende, Nucl. Phys. B \textbf{303},
407 (1988).

\bibitem{maggiore1993} M. Maggiore, Phys. Lett. B \textbf{304}, 65
(1993).

\bibitem{Ng1994} Y. J. Ng and H. van Dam, Mod. Phys. Lett. A \textbf{9},
335 (1994).

\bibitem{kempf1995} A. Kempf, G. Mangano, R. B. Mann, Phys. Rev.
D \textbf{52}, 1108 (1995).

\bibitem{Rovelli1996} C. Rovelli, Phys. Rev. Lett. \textbf{77}, 3288
(1996).

\bibitem{scardigli1999} F. Scardigli, Phys. Lett. B \textbf{452},
39 (1999).

\bibitem{adler1999} R. J. Adler, D. I. Santiago, Mod. Phys. Lett.
A \textbf{14}, 1371 (1999).

\bibitem{tawfik2015} A. N. Tawfik, A. M. Diab, Rep. Prog. Phys. \textbf{78},
126001 (2015).

\bibitem{scardigli2017} F. Scardigli, G. Lambiase, E. C. Vagenas,
Phys. Lett. B \textbf{767}, 242--246 (2017).

\bibitem{benedetto2015} E. Benedetto, A. Feoli, Mod. Phys. Lett.
A \textbf{30}, 1550075 (2015).

\bibitem{luciano2019} G. G. Luciano, L. Petruzziello, Eur. Phys.
J. C \textbf{79}, 283 (2019).

\bibitem{chung2019} W. S. Chung, H. Hassanabadi, Phys. Lett. B\textbf{
793}, 451--456 (2019).

\bibitem{wald1984} R. M. Wald, \textit{General Relativity}, (The
University of Chicago Press, Chicago, 1984).

\bibitem{jackiw2003} R. Jackiw, S. Y. Pi, Phys. Rev. D \textbf{68},
104012 (2003).

\bibitem{gosh2010} S. Gosh, arXiv:1003.0285v3 {[}hep-th{]}.

\bibitem{nozari2012} K. Nozari, P. Pedram, M. Molkara, Int. J. Theor.
Phys. \textbf{51}, 1268--1275 (2012).

\bibitem{santos2012} M. A. Santos, I. V. Vancea, Mod. Phys. Lett.
A \textbf{27}, 1250012 (2012).

\bibitem{moradpour2018A} H. Moradpour, A. Sheykhi, C. Corda, I. G.
Salako, Phys. Lett. B \textbf{783}, 82--85 (2018).

\bibitem{bagchi2019} B. Bagchi, A. Fring, Int. J. Mod. Phys. B \textbf{33},
1950018 (2019).

\bibitem{pikhitsa2010} P. V. Pikhitsa, arXiv:1010.0318v3 {[}astro-ph.CO{]}.

\bibitem{zhe2011} C. Zhe, L. Ming-Hua, L. Xin, Commun. Theor. Phys.
\textbf{56}, 184--192 (2011).

\bibitem{frieman2008} J. Frieman, M. Turner, D. Huterer, Ann. Rev.
Astron. Astrophys. \textbf{46}, 385 (2008).

\bibitem{padmanabhan2009} T. Padmanabhan, Adv. Sci. Lett. \textbf{2},
174 (2009).

\bibitem{felice2010} A. De Felice, S. Tsujikawa, Living Rev. Relativity
\textbf{13}, 3 (2010).

\bibitem{randall1999} L. Randall, R. Sundrum, Phys. Rev. Lett. \textbf{83},
4690 (1999).

\bibitem{awad2014} A. Awad, A. F. Ali, Cent. Eur. J. Phys. \textbf{12},
245-255 (2014).

\bibitem{lambiase2018} G. Lambiase, F. Scardigli, Phys. Rev. D \textbf{97},
075003 (2018).
\end{thebibliography}
\end{document}